# Spontaneous formation of optically induced surface relief gratings

H. Leblond[1], R. Barille[1], S. Ahamadi-kandjani[1,2], J.-M. Nunzi[1,3], E. Ortyl,[4] , and S. Kucharski[4]

[1]Laboratoire POMA, Université d'Angers, FRE CNRS 2899
2, Bd Lavoisier, 49045 Angers, France
[4] Wroclaw University of Technology, Faculty of Chemistry,
Department of Polymer Engineering and Technology,
50-370 Wroclaw, Poland.

**Abstract**:
We develop a model based on Fick's law of diffusion as a phenomenological description of the molecular motion, and on the coupled mode theory, to describe single-beam surface relief grating formation in azopolymers thin films. The model allows to explain the mechanism of spontaneous patterning, and self-organization. It allows to compute the surface relief profile and its evolution with good agreement with experiments.

## 1. Introduction

The use of tran-cis isomerization of the azo moiety is still a topic of research and offers large possibilities of new developments. Until recently the assembly of artificial molecular machines allowing precisely controlled motion that responds to light has been demonstrated [1 - 2]. Beyond these activities the azobenzene chromophore continues to present new and unique optical effect especially for the formation of surface patterns in azopolymer thin films. This chromophore has also served as an photochemical model of cognitive ability [3]. From the applicative point of view, polymer thin films allow a large number of optoelectronic devices [4]. The most interesting property of the azobenzene molecules is the efficient and reversible photochemical isomerization which occurs between the trans and cis geometric isomers. Isomerization involves rotation or inversion around the azo bond. The formation of surface topography in the azo systems requires considerable material motions over micron length scales. Possible mechanisms for the process explaining single step room temperature all-optical patterning of thin films haves been proposed. They try to explain the origin of the driving force responsible for Surface Relief Grating (SRG) inscription. Beyond them we note diffuse mechanisms [5], pressure gradients, etc… More recently a random walk model has been proposed [6], it was introduced previously by basic principles for an engine transforming energy into directed translational or rotational motion at the nanoscale [7]. All these interpretations induced much controversy and none of the proposed mechanisms appears to provide a suitable explanation for all observations, especially in the case of multistate polarization addressing using a single beam in an azopolymer film.

[2] Present address: RIAPA, University of Tabriz, 51664 Tabriz, (Iran).
[3] Present address: Department of Chemistry, Chernoff Hall, room 310, Queen's University, 90 Bader Lane, Kingston ON, Canada K7L 3N6.

A SRG can indeed form in the film through illumination by a single beam. Since in this case no interference pattern is applied, some self-organization mechanism is involved. The aim of this paper is to explain this mechanism. The detail of the microscopic mechanism responsible for the molecular motion of the azopolymer is very complicated but influences only slightly the self-organization process. Hence, we will describe it by means of a phenomenological model, whose simplicity would permit the treatment of the quite complicated nonlinear optical interactions involved by the self-organization process. This model is based on Fick's law of diffusion that scales with light intensity. Further, light is trapped in the sample and couples into counterpropagating guided modes, which interact with the azopolymer. Using this approach, we can explain the self-patterning of the azopolymer film lightened with a single beam, and fit our measurements.

## 2. The model

### 2.1. The diffusion equation

Ficks's law of diffusion reads as

$$\Phi = -KS\nabla c\,, \tag{1}$$

where $\Phi$ is the particle flow, $c$ the concentration, $S$ the diffusing surface, $\nabla$ the gradient operator and $K$ some constant. Here the concentration is that of moving molecules, which are the cis-isomers of the azo-dye. It is proportional to both the quantity $\rho$ of azo-dyes by surface unit of the film, and the light intensity $I$. We have thus

$$c = qI\rho\,, \tag{2}$$

where $q$ is some constant. It is well-known that the diffusion is anisotropic, and depends on the polarization direction. An anisotropic version of (1) is

$$\Phi = -KSA\nabla c\,, \tag{3}$$

where $A$ is some tensor depending on the polarization direction of the light. For sake of simplicity and according to the experiment, we assume that the light is linearly polarized along the direction $x$, and that

$$A = \begin{pmatrix} A_{xx} & 0 & 0 \\ 0 & 0 & 0 \\ 0 & 0 & 0 \end{pmatrix} \quad (4)$$

is a projector onto this direction of the incident light, to give account for diffusion in the polarization direction only. In this case, the flow $\Phi$ has only one component along $x$, which is

$$\Phi = -\alpha \frac{\partial (I\rho)}{\partial x}, \quad (5)$$

where we have set $\alpha = KqA_{xx}S$. The model becomes thus one-dimensional, $x$ being the direction in the film plane along which the intensity varies. We will disregard below the variations of the diffusion surface $S$, although the film thickness is not uniform.

The conservation of matter expresses as

$$\frac{\partial \rho}{\partial t} + \nabla \cdot \Phi = 0, \quad (6)$$

which reduces in (1+1) dimensions, using (4), to

$$\frac{\partial \rho}{\partial t} = \alpha \frac{\partial^2 (I\rho)}{\partial x^2} \ . \quad (7)$$

### 1.2. The model accounts for surface relief grating formation

Before we proceed to the analysis of self-patterning, let us show that the phenomenological model yielded by Eq. (7) is able to describe SRG formation when a definite interference pattern is applied. We assume a periodical light intensity $I = I_0 \cos^2 kx,$ and look for stationary solutions of Eq. (7). Then $I\rho$ must be some constant $b$, thus

$$\rho = \frac{b}{I_0 \cos^2 kx} \ . \quad (8)$$

Since the film thickness is proportional to $\rho$, equation (8) describes a SRG. The thickness is maximal in the dark, as in experiments. Obviously this phenomenological model is very simplified, and does not account for the complex microscopic dynamics of the grating formation. Also, the model does not contain any term which can limit the growth of the grating, therefore according to relation (8) the density $\rho$ is infinite as $x = (n + 1/2)\pi / k$. This can be avoided by adding a spontaneous diffusion flow

$$\Phi_s = -K' S \nabla \rho \quad (9)$$

to the model. In (1+1) dimensions, equation (7) becomes

$$\frac{\partial\rho}{\partial t}=\alpha\frac{\partial^2\big((I+a)\rho\big)}{\partial x^2}\ , \tag{10}$$

where we have set $a=K'S/\alpha$. This way Eq. (8) becomes

$$\rho=\frac{b}{I_0\cos^2 kx+a}\ , \tag{11}$$

and the divergence is suppressed. Notice that a spontaneous diffusion would lead to the instability of the grating when the light is off, which is not the case in experiments.

### 2.3. Equation for the coupled modes

The refractive index of the film (1.48) is close to the one of the substrate (1.52), and lower. Light can be guided in the substrate, but also in the whole structure constituted by both the film and the substrate, the guiding conditions being much more easily satisfied in the latter case. Due to Rayleigh-type diffusion on impurities and surface defects, a tiny part of the incident light, normal to the plane of the sample, is coupled into it. Let us denote the electric field of the incident light (above the film) by

$$E=T(x)e^{ik_v z}\ , \tag{12}$$

$k_v$ being the wave vector of the incident light in vacuum, and the light coupled into the substrate and film, propagating in both directions, by

$$R(x)e^{-ik_0 x}\ \text{ and }\ S(x)e^{+ik_0 x}\ . \tag{13}$$

(cf. Fig. 1). Notice that the active part of these waves is the part which propagates in the film and not in the substrate. At the beginning of the process, the amplitudes $R$ and $S$ are very small. They form some interference pattern, which starts to induce a SRG. Some component of this induced grating is adequately matched and produces a coupling between the two waves $R(x)e^{-ik_0 x}$ and $S(x)e^{+ik_0 x}$. Then the amplitudes $R$ and $S$ grow, and the induced grating also does. This increases the coupling of the incident wave into the sample, and finally a standing wave with appreciable amplitude may form in it, inducing in the meanwhile a regular grating with appreciable amplitude. The analysis below is intended to prove that this process actually occurs. It is based on the coupled mode theory [8]. Inside the sample and within the effective index approximation, the wave equation is

$$\Delta E+k^2E=0\,, \tag{14}$$

where $\Delta$ is the Laplacian operator, and $k$ the propagation constant of the guided mode. We write the total electric field $E$ inside the sample as

$$E=T(x)+R(x)e^{-ik_0 x}+S(x)e^{ik_0 x}\ , \tag{15}$$

where $R$, $T$, and $S$ vary slowly with the variable $x$ along the sample (Fig. 1) and $T$ accounts for the incident pump beam, and $R$ and $S$ for the light propagating in both directions along the plane of the sample. $k_0$ is a reference value for the propagation constant $k$. With respect to [8], we have added the fist term which accounts for the incident wave.

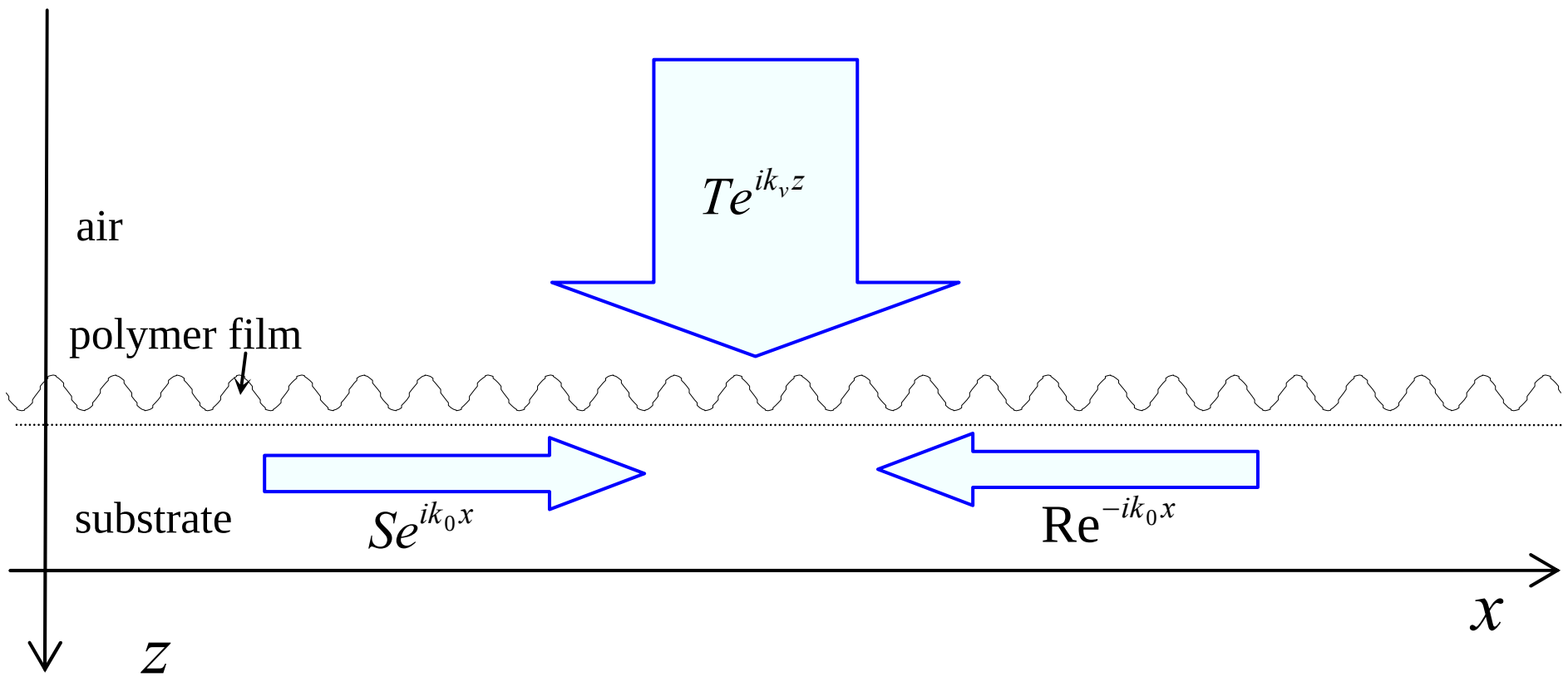

Figure 1: Schematic section of the sample, showing the different waves involved in the theoretical analysis.

The effective propagation constant $k$ depends on the film thickness (thus on the surface density $\rho$ of section II.A) trough the modal dispersion relation of the planar guide. Since the exact dependency $k(\rho)$ is too complicated to be tractable, we admit that $k$ can be linearized about the initial value $\rho_0$ of the density $\rho$, as

$$k(\rho_0 + \delta\rho) \approx k(\rho_0) + \frac{dk}{d\rho}(\rho_0)\delta\rho\,, \tag{16}$$

thus

$$k^2 = k_0^2 + \gamma\delta\rho\,, \tag{17}$$

where $k_0 = k(\rho_0)$ is the reference value of the propagation constant used in (15). Notice that, since the guiding layer is composed of both the substrate and the film, the variations of $k$ remain small when the modulation depth of the grating reaches the entire film thickness. We report Eqs. (15) and (17) into (14), consider $\delta\rho$ and the first-order derivatives as small quantities of the same order, and neglect the second order derivatives of $R$, $S$, and $T$, to obtain

$$-2ik_0\frac{\partial R}{\partial x}e^{-ik_0 x} + 2ik_0\frac{\partial S}{\partial x}e^{ik_0 x} + \gamma\delta\rho\left(T + Re^{-ik_0 x} + Se^{ik_0 x}\right) = 0\,. \tag{18}$$

Notice that the exact nature of the coupled modes is ignored, as in [8]. Further, since the guided waves propagate in both the film and substrate (which is obviously thick with respect to the wavelength), the number of modes is huge, and it is likely that several but not all of them are effectively involved in the process. We disregard this problem, and consider a single mode in each direction, for sake of simplicity. In an analogous way, due to Fresnel reflections at the interfaces, the incident beam forms a standing wave in the $z$ direction, which has been replaced by $T$ , independent of $z$ .

## 3. Fourier analysis and spontaneous patterning

The spontaneous arising of a pattern, related to the growth of counterpropagating waves (13) can be demonstrated from Eqs. (10) and (18) above, by means of a perturbative approach. This is the aim of the present section.

### 3.1. Fourier components of the model equation

Let us first consider the intensity of the light propagating inside the sample, which is $I = |E|^2$ , where $E$ is given by equation (15). It is convenient to write $I$ as

$$I = \sum_{n=-2}^{2} I_n e^{ink_0 x} , \tag{19}$$

with

$$I_0 = |T|^2 + |R|^2 + |S|^2 , \quad I_1 = I_{-1}^* = TR^* + T^* S , \quad I_2 = I_{-2}^* = R^* S . \tag{20}$$

The grating pitch $\Lambda$ is expected to coincide with the spatial period of $I$ , as in the case of a two-beam experiment. Hence, according to Eq. (19), we should have $\Lambda = 2\pi / k_0$ . According to the form of the equation and to the expansion of $I$ , we expand $\delta\rho$ in a Fourier series of the propagation constant $k_0$ , as

$$\delta\rho = \frac{2k_0}{\gamma} \sum_{n=-\infty}^{+\infty} A_n e^{ink_0 x} . \tag{21}$$

Substitution of (21) into equation (18) yields the following set of equations:

$$(n=1) \qquad i\frac{\partial S}{\partial x} + A_1 T + A_2 R + A_0 S = 0 , \tag{22}$$

$$(n=-1) \qquad -i\frac{\partial R}{\partial x} + A_{-1} T + A_0 R + A_{-2} S = 0 , \tag{23}$$

$$(n \neq \pm 1) \qquad A_n T + A_{n+1} R + A_{n-1} S = 0 . \tag{24}$$

Further, we report the Fourier expansions (19) and (21) into the diffusion equation (10), which yields

$$\frac{\partial A_n}{\partial t} = \alpha \frac{\partial^2}{\partial x^2} \sum_{m+p=n} \left( I_m + a\delta_{m,0} \right) \left( A_p + \frac{\gamma}{2k_0} \rho_0 \delta_{p,0} \right) \tag{25}$$

( $\delta_{p,0}$ is the Kronecker symbol). Using (20), it reduces to

$$\frac{\partial A_n^{'}}{\partial t} = \alpha \frac{\partial^2}{\partial x^2} \left[ \left( a + |T|^2 + |R|^2 + |S|^2 \right) A_n^{'} + \left( TR^* + T^* S \right) A_{n-1}^{'} + R^* S A_{n-2}^{'} + \left( T^* R + TS^* \right) A_{n+1}^{'} + RS^* A_{n+2}^{'} \right]$$

(26)

where we have set

$$A_n^{'} = A_n + \frac{\gamma}{2k_0} \rho_0 \delta_{n,0} \, . \tag{27}$$

Eqs. (22-24,26) give the evolution of the Fourier modes $A_n$ of the density $\rho$ of azo-dye, which measures the film thickness in our approach.

### 3.2. Instability of the flat state

The above equations obviously admit the flat state solution $R = S = 0$, $\delta\rho = 0$. In order to prove that a grating forms spontaneously, we will show that this state is unstable (by linear stability analysis). We assume a uniform lightning, $T$ real given, and solve equations (22-24,26) in a perturbative way. At the beginning of the process, $R$ and $S$ are small. Therefore we introduce a small parameter $\varepsilon$, replace the unknowns $R$ and $S$ by $\varepsilon R$ and $\varepsilon S$ respectively, and expand the $A_n$ in power series of $\varepsilon$, as $A_n = A_n^0 + \varepsilon A_n^1 + \ldots$. At order $\varepsilon^0$, all the $A_n^0$ with $n \neq 0$ are 0, and $A_0^{'0} = \gamma\rho_0 / (2k_0) = b$ for shortening. At order $\varepsilon^1$, the only nonzero terms are $A_1^1$ and $A_{-1}^1 = A_1^{1*}$, which satisfy

$$\frac{\partial A_1^1}{\partial t} = \alpha \frac{\partial^2}{\partial x^2} \left[ \left( a + T^2 \right) A_1^1 + bT \left( R^* + S \right) \right]. \tag{28}$$

The equations for $R$ and $S$ drawn from equations (22-23) are

$$i\frac{\partial S}{\partial x} + A_1^1 T = 0 \, , \qquad -i\frac{\partial R}{\partial x} + A_1^{1*} T = 0 \, . \tag{29}$$

We introduce some function $\varphi$ satisfying $A_1^1 = \partial\varphi / \partial x$, so that Eqs. (28) reduce to $S = iT\varphi$, $R = -iT\varphi^*$ .The equation governing the evolution of $\varphi$ is then

$$\frac{\partial \varphi}{\partial t} = \alpha \frac{\partial}{\partial x} \left[ \left( a + T^2 \right) \frac{\partial \varphi}{\partial x} + 2ibT^2 \varphi \right]. \tag{30}$$

We look for solutions of equation (30) of the form $\varphi = g e^{\lambda t + i\kappa x}$, and compute the growth rate

$$\lambda = -\alpha\kappa \left[ \left( a + T^2 \right) \kappa + 2bT^2 \right] \, . \tag{31}$$

Hence $\lambda$ is positive for $-2bT^2 / \left( a + T^2 \right) < \kappa < 0$, which proves that the solution with $R \equiv S \equiv 0$ is unstable. Therefore nonzero $R$, $S$, and $A_1^1$ will arise spontaneously. This shows that the interaction between coupled waves and molecular motion is able to induce the spontaneous symmetry breaking required for the self-organization of the surface yielding the SRG.

## 4. Numerical study

### 4.1. Numerical procedure

We solve numerically equations (10) and (14), in which $k^2$ is given by (17). The incident light $Te^{ik_0 z}$ is taken into account by adjoining to equation (14) the boundary condition $E \xrightarrow[x\to\pm\infty]{} T$ (or alternatively by adding the constant $T$ to the solution of (14) with the boundary condition $E \xrightarrow[x\to\pm\infty]{} 0$). The evolution equation (10) is solved by means of a modified Euler scheme, the $x$-derivatives being computed by three points finite differences. The wave equation (14) is discretized by means of three points finite differences. The resulting linear system is solved at each half time step using the Gauss algorithm.

In the simulation we control the incident intensity by adjusting the parameter $T$. The parameter $\gamma$ which accounts for the variation rate of the wave vector with respect to the film thickness measured by the surface density $\rho$ is also adjusted. The number of points in the $x$ direction and the time step are adjusted to ensure the convergence of the numerical scheme.

### 4.2. Evolution of the SRG

The samples are polymer films made from a highly photoactive azobenzene derivative containing heterocyclic sulfonamide moieties: 3-[{4-[(E)-(4-{[(2,6-dimethylpyrimidin-4-yl) amino] sulfonyl}phenyl) diazenyl]phenyl}-(methyl)amino]propyl 2-methylacrylate [9]. Thin films on glass substrates were prepared by spin-coating of the polymer from THF solutions with a concentration of 50 mg/ml. The thickness measured with a Dektak-6M Stylus Profiler was around 1μm. Absorbance at λ = 438 nm maximum is 1.9 . The λ = 476.5 nm laser line of a continuous argon ion laser is used to excite the azo polymer absorption close to its absorption maximum. The absorbance at working wavelength is 1.6. The incoming light intensity is controlled by the power supply. The polarization direction of the laser beam is varied using a half-wave plate. The sample is set perpendicular to the incident laser beam. The size of the collimated laser beam impinging onto the polymer sample is controlled with a Kepler-type afocal system. The sample is irradiated with different polarizations using different laser beam intensities with a defect beam size of 4 mm diameter at $1/e^2$ .

In a preliminary experiment we carefully checked that laser irradiation leads to a topographic modification of the polymer surface resulting in a SRG. When the surface relief is formed, the impinging beam itself is diffracted in several diffraction orders. We show in Fig. 2 the recorded intensity of the first diffraction order as a function of time, for different laser beam intensities. The diffraction which occurs in both forward and backward directions is collected in the backward direction by a f = 200 mm focal-length lens and registered as a function of time by a photodiode. Above 760 mW/cm$^2$ input beam intensity, the polymer film could be damaged and we did not exceed this limit. Within the Raman-Nath approximation, which is valid here, as shows the symmetry of the diffraction pattern, the diffracted intensity is proportional to the square of the amplitude of the SRG modulation [10], hence to the squared amplitude of the oscillations of $\delta\rho$, in our model. The quantity $(\max_x \delta\rho)^2$, with adequate normalization, is plotted vs time on Fig. 2. Numerical computation shows a very good agreement with experimental results. In the simulation we vary the parameter $T$ to change the incident intensity. The first step has been to calibrate the intensity in the simulation with different experimental results. The self-organized SRG formation phenomenon exhibits a threshold depending on the power density. Simulations and experimental results show the same threshold value. We define a threshold time for the SRG formation by means of the second derivative threshold calculation method which locates the point of maximum rate of change of the curve. The threshold evolves linearly as a function of the input beam intensity and its slope is estimated at about 33 mW/cm$^2$.mn$^{-1}$. We checked that this threshold does not depend on polarization. Measurements using different beam sizes between 3 and 6 mm confirmed also that the threshold time is a function of power density only. We note that in our experiment the grating wave-vector is parallel to the linear polarization of the laser beam, in accordance with the theory.

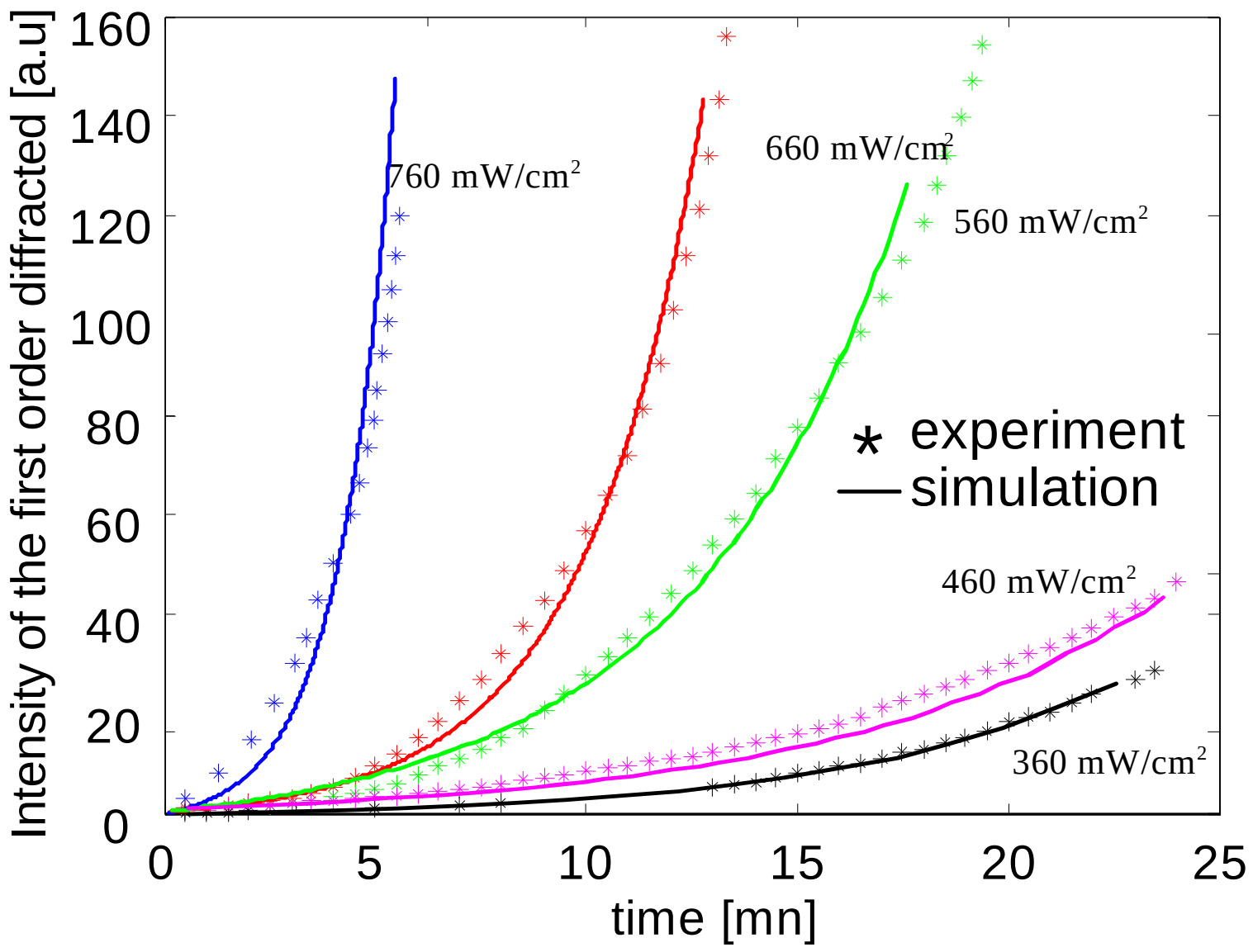

Figure 2: Intensity of the first diffraction order measured as a function of time for different laser beam intensities and comparison with the simulation. The latter is the SRG squared amplitude $(\max_x \delta\rho)^2$ computed numerically and adequately normalized.

Fig. 3 presents the complete measurement of the intensity diffraction from the threshold to the saturation level. The beam power required to create the SRG is 100 mW and the intensity used is 140 mW/cm$^2$. In that case the experimental measurements fit the simulation and show a saturation in the diffraction intensity after one hour. The parameters for the simulations are the same as used previously for the same beam power (the simulation is the same as Fig. 2, but run longer).

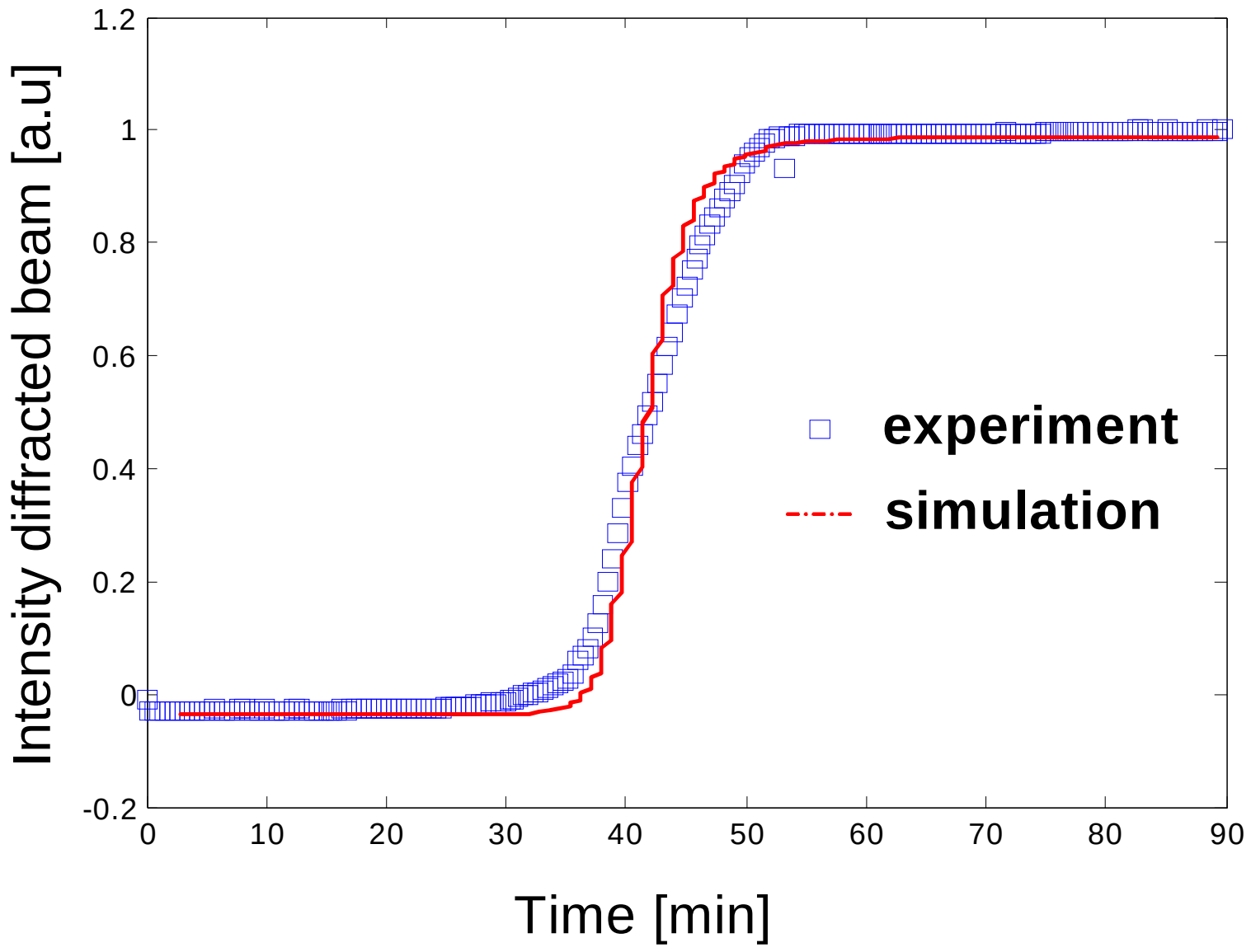

Figure 3: Evolution of intensity of the first diffraction order measured as a function of time for a beam intensity of 450 mW/cm$^2$. The saturation level of the recorded grating is reached in 50 min. The theoretical curve is defined as in Fig. 2.

### 4.3. SRG profiles

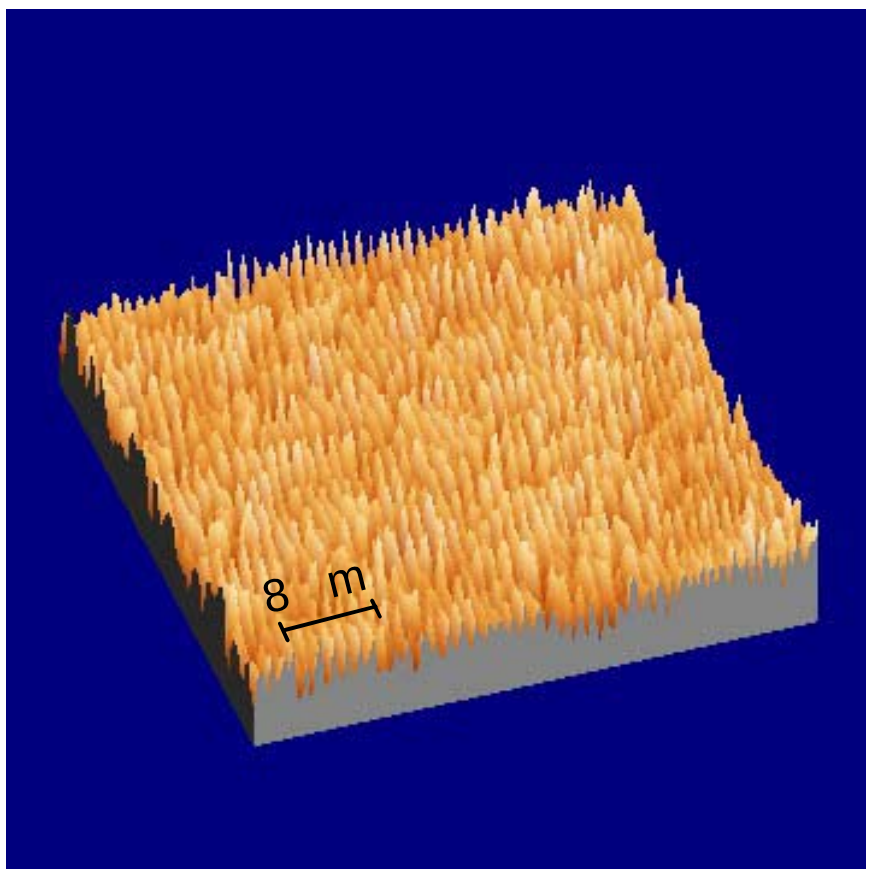

Figure 4: AFM image of a typical surface grating displaying a SRG amplitude of 50 nm The grating is made with an Argon laser with wavelength of 476 nm and with intensity of 450 mW/cm$^2$. The material is the same as in [9].

Fig. 4 shows an AFM (atomic force microscopy) topographic images of a permanent structure induced with the input laser beam. The laser beam intensity was 450 mW/cm$^2$. The growth time for this grating was one hour. The SRG has a depth of 50 nm ± 5 nm. From this topographic measurement it is possible to determine the surface profile. It is given in Fig. 5. From this figure the grating pitch is calculated and an average value is $\Lambda = 800\ \mathrm{nm} \pm 30\ \mathrm{nm}$.

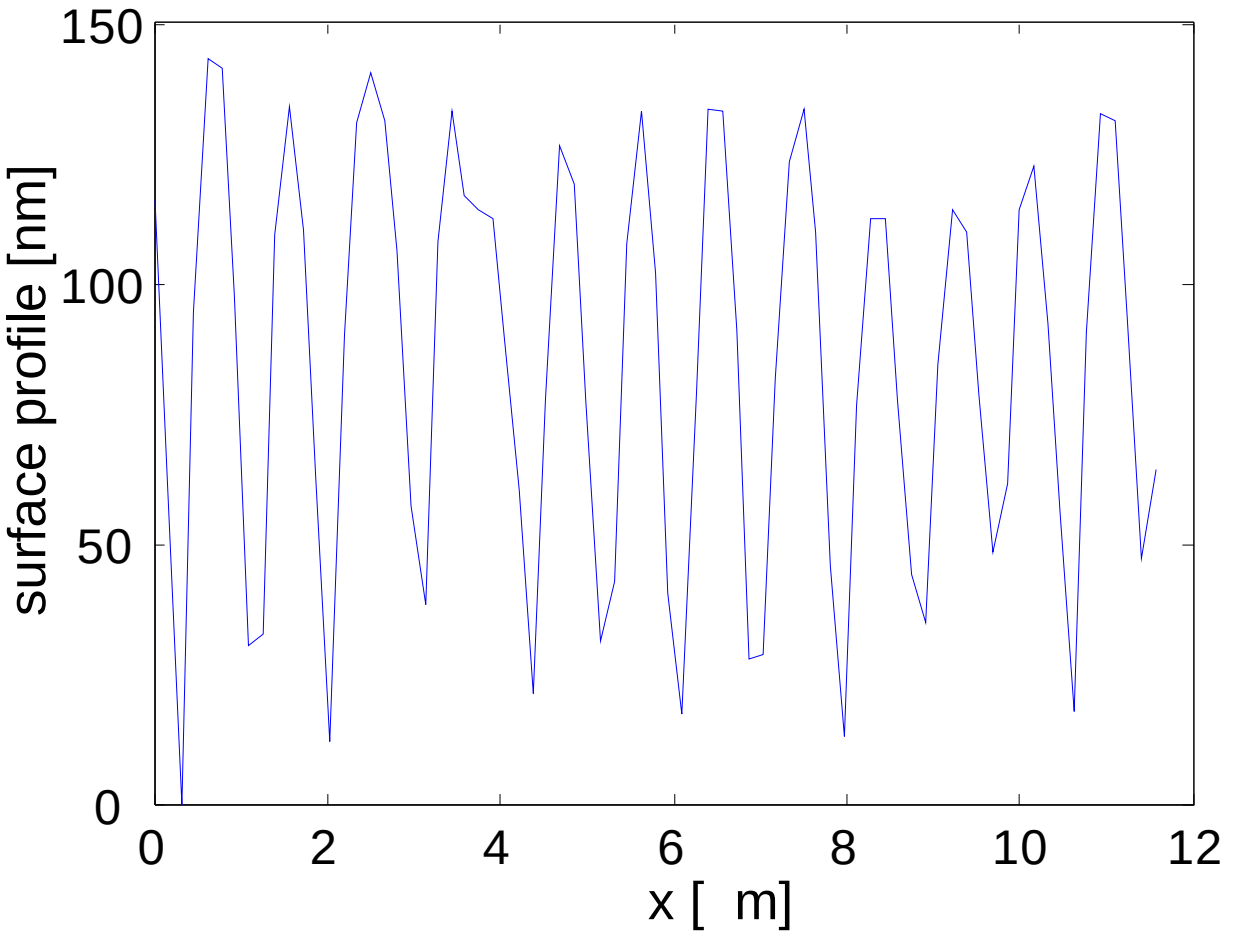

Figure 5: Topographic surface profile obtained with an AFM for the SRG shown in Fig. 4. The grating pitch is calculated and its an average value is $\Lambda = 800\ \mathrm{nm} \pm 30\ \mathrm{nm}$.

In the simulation as in the experiment the SRG results from isomerization induced translation in which the molecules migrate almost parallel to the polarization direction. When the collimated laser beam with intensity $I = T^2$ impinges onto the sample, light is diffused inside the sample in all directions around any micro-roughness. The surface profile at the very beginning of the simulation is shown in Fig. 6a. This small roughness will couple a small amount of ~~a~~ light into the plane of the sample. The propagation of this light inside the sample is taken into account by Eq. (15). The roughness height increases with time $t$ from its initial value up to saturation. Indeed, the process which initially diverges will finally saturate, when the depth of the SRG modulation attains the entire thickness of the film. Fig. 6b shows the same surface profile when the saturation value is reached. A perfect grating is obtained on the surface. The pitch obtained numerically is $\Lambda = 2\pi / k_0$, as expected from the analytic part of the theory. The value of the reference propagation constant $k_0$ used in equation (17) in the simulation is adjusted to give the correct grating pitch. The effective index $n_{eff} = k_v / k_0$ can then be estimated to $n_{eff} \approx 1.68$. It can be written as $n_{eff} = n / \cos\theta$, with $n \approx 1.52$ the refractive index of the substrate and $\theta$ the angle between the optical ray and the $x$-axis, within the geometrical optics approximation. We obtain $\theta \approx 25°$, which can be compared to the limit angle for total reflection on the substrate-air interface, which corresponds to $\theta \approx 49°$. This value confirms that the guided modes involved in the process propagate between the two external surfaces of the sample, in contact with air. The selection of the modes corresponding to this value of $n_{eff}$, in the middle of the waveguide spectrum, is not explained at the present state of the theory. The instability rate of the various modes, the intensity of their coupling with the incident waves, or the interferences between the modes, may be involved in this process, which is left for further investigation. Notice that the final profile is close to the absolute value of a sine function.

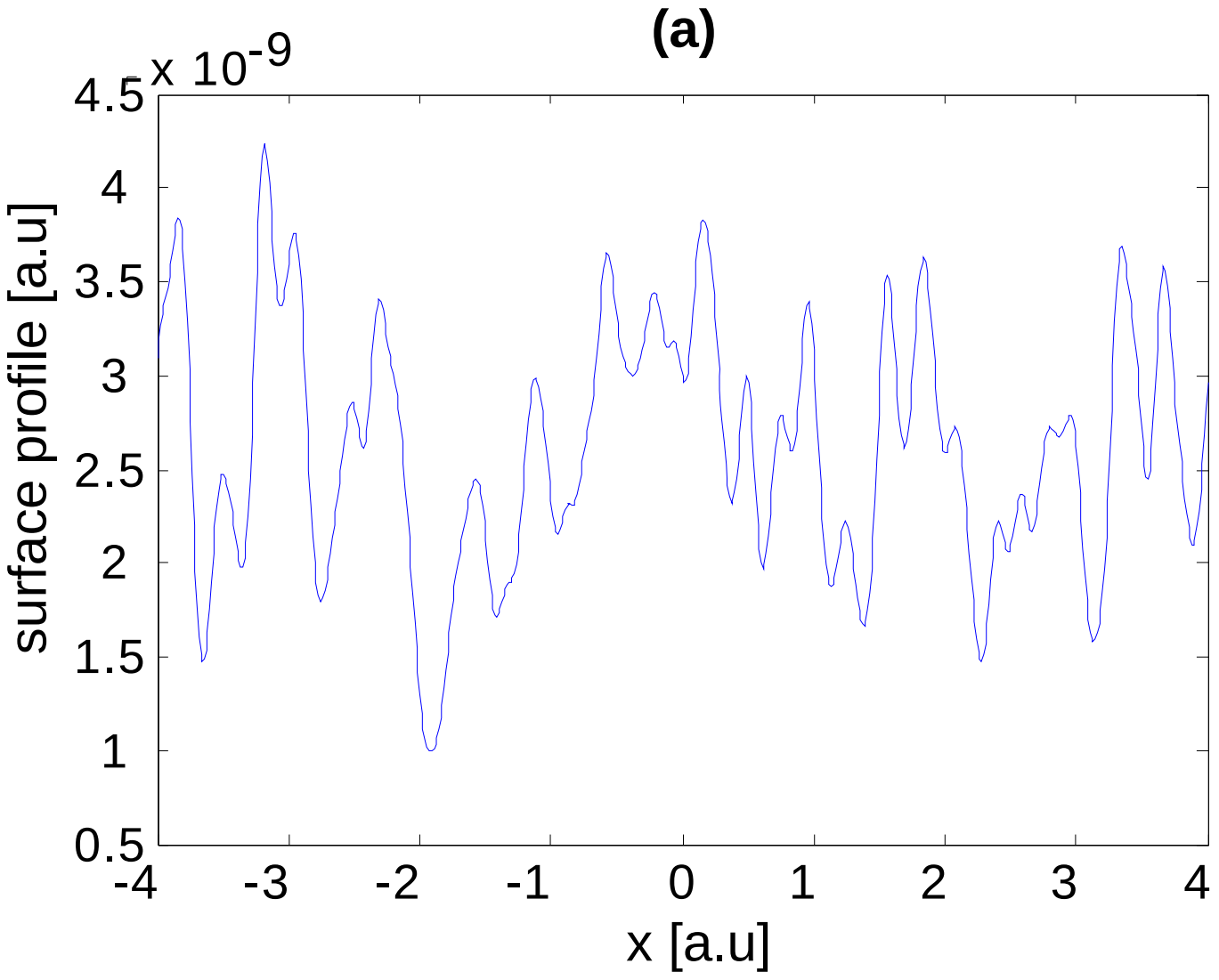

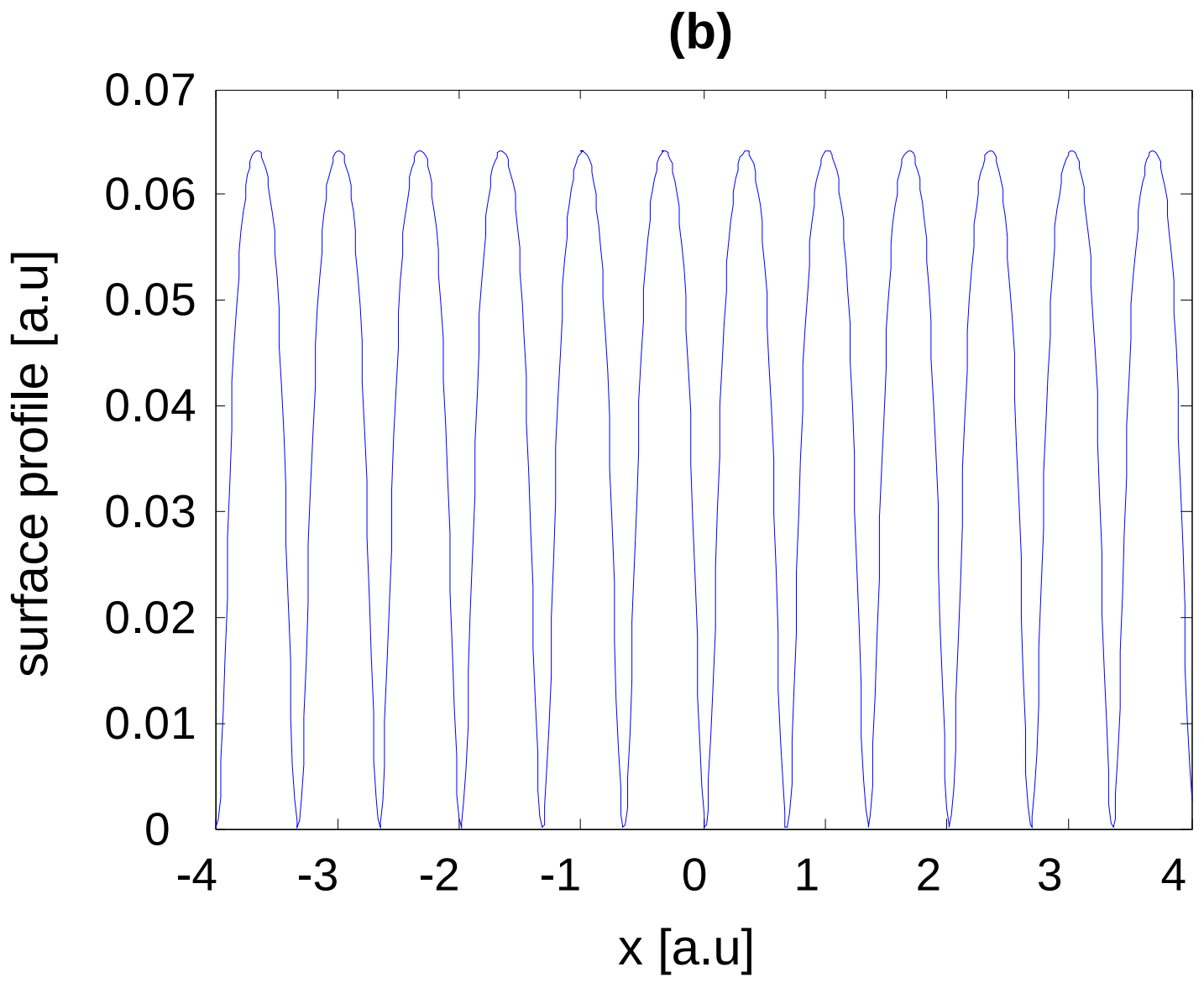

Figure 6: Transverse surface profile (i.e. $\delta\rho$ ) at the beginning of our simulation (a). The initial data consider a random surface that characterizes the surface roughness. Comparison between the transverse surface profile evolution at the beginning of the simulation (a) and at the end of the simulation (b) when the saturation level is reached.

## 5. Conclusion

In conclusion, the surface profile and intensity evolution of azopolymer thin films during exposition to a single laser beam is explained by means of a model based on both the coupled mode theory and a phenomenological description of the molecular motion. The SRG results from a reversible trans-cis-trans isomerization process lead to a quasi-permanent molecular reorientation in a direction perpendicular to the light polarization directions. The models reported in the literature referred to experiments to create patterns on the surface of azopolymer thin films using the exposition to interference patterns of light at a wavelength of absorption band, and were intended to explain the pattern formation at the molecular level. Here we have explained the mechanism of spontaneous patterning which can be obtained in a single-beam experiment. Therefore we use a phenomenological model based on Ficks's law of diffusion to account for the molecular motion. The process involves the propagation of waves guided in the sample (film and substrate), which is treated as a planar waveguide in the effective index approximation. The wavelength of the incident pump beam, together with the guiding properties, characterizes the optical frequency of light propagating inside the sample. This frequency propagates in the sample with a given propagation constant and wavelength: our analysis shows that this wavelength coincides with the grating pitch. Good agreements between theory and experiments are found, which will allow us to extrapolate further mechanisms of pattern formation.

**Figure captions:**

Figure 1: Schematic section of the sample, showing the different waves involved in the theoretical analysis.

Figure 2: Intensity of the first diffraction order measured as a function of time for different laser beam intensities and comparison with the simulation. The latter is the SRG squared amplitude $\left(\max_x \delta\rho\right)^2$ computed numerically and adequately normalized.

Figure 3: Evolution of intensity of the first diffraction order measured as a function of time for a beam intensity of 450 mW/cm$^2$. The saturation level of the recorded grating is reached in 50 min. The theoretical curve is defined as in Fig. 2.

Figure 4: AFM image of a typical surface grating displaying a SRG amplitude of 50 nm The grating is made with an Argon laser with wavelength of 476 nm and with intensity of 450 mW/cm$^2$. The material is the same as in [9].

Figure 5: Topographic surface profile obtained with an AFM for the SRG shown in Fig. 4. The grating pitch is calculated and its an average value is $\Lambda = 800\text{ nm} \pm 30\text{ nm}$.

Figure 6: Transverse surface profile (i.e. $\delta\rho$) at the beginning of our simulation (a). The initial data consider a random surface that characterizes the surface roughness. Comparison between the transverse surface profile evolution at the beginning of the simulation (a) and at the end of the simulation (b) when the saturation level is reached.